# The efficiency of multi-target drugs: the network approach might help drug design

Péter Csermely[1*], Vilmos Ágoston[2] and Sándor Pongor[2,3]

[1]Department of Medical Chemistry, Semmelweis University, P O Box 260., H-1444 Budapest 8, Hungary
[2]Szeged Biological Research Center, P O Box 521. H-6701 Szeged, Hungary
[3]International Centre for Genetic Engineering and Biotechnology (ICGEB), Padriciano 99, I-34012 Trieste, Italy

**Despite considerable progress in genome- and proteome-based high-throughput screening methods and rational drug design, the number of successful single target drugs did not increase appreciably during the past decade. Network models suggest that partial inhibition of a surprisingly small number of targets can be more efficient than the complete inhibition of a single target. This and the success stories of multi-target drugs and combinatorial therapies led us to suggest that systematic drug design strategies should be directed against multiple targets. We propose that the final effect of partial, but multiple drug actions might often surpass that of complete drug action at a single target. The future success of this novel drug design paradigm will depend not only on a new generation of computer models to identify the correct multiple hits and their multi-fitting, low-affinity drug candidates but also on more efficient *in vivo* testing.**

**Multi-target drugs help us more often than we think**
Drug development strategies have been influenced profoundly by the wealth of potential targets offered by genome projects. At present, the goal is to: (i) find a target of suitable function; (ii) identify the 'best-binder' by high-throughput screening of large combinatorial libraries and/or by rational drug design based on the three-dimensional structure of the target; (iii) provide a set of proof-of-principle experiments; and (iv) develop a technology platform projecting to potential clinical applications (Figure 1a). However, despite all the careful studies and the considerable drug-development efforts undertaken, the number of successful drugs and novel targets did not increase appreciably during the past decade [1,2]. Several highly efficient drugs, such as non-steroidal anti-inflammatory drugs (NSAIDs), salicylate, metformin or Gleevec™, affect many targets simultaneously. Furthermore, combinatorial therapy, which represents another form of multi-target drugs, is used increasingly to treat many types of diseases, such as AIDS, cancer and atherosclerosis [3-5]. Snake and spider venoms are both multi-component systems and plants also employ batteries of various factors to fence off pathogenic attack; thus the use of multiple molecules is apparently an evolutionary success story. Finally, traditional medical treatments often use multi-component extracts of natural products. Based on these examples and on our recent results of network analysis [6] here we propose that systematic drug design strategies should be directed against multiple targets, and this novel drug design paradigm might often result in the development of more efficient molecules than the currently favored single-target drugs.

**Single hits are often insufficient**
Agents that affect one target only ('single-hits') might not always affect complex systems in the desired way even if they completely change the behavior of their immediate target. For example, single targets might have 'back-up' systems that are sometimes different enough not to respond to the same drug, and many cellular networks are robust and prevent major changes in their outputs despite dramatic changes in their constituents [7,8]. These considerations are independent of whether or not the pharmacological agent inhibits or activates its target.

*Corresponding author:* Peter Csermely (csermely@puskin.sote.hu; www.weaklink.sote.hu).

## Multi-target drugs are often low-affinity binders

Development of a multi-target drug is likely to produce a drug that interacts with lower affinity than a single target drug because it is unlikely that a small, drug-like molecule will bind to a variety of different targets with equally high affinity. However, low-affinity drug binding is apparently not a disadvantage. For example, memantine (a drug used to treat Alzheimer's disease) and other multi-target non-competitive NMDA receptor antagonists show that low-affinity, multi-target drugs might have a lower prevalence and a reduced range of side-effects than high-affinity, single-target drugs [9,10]. Does low-affinity binding mean that the interaction of the drug with the target is non-productive? Not necessarily. Most components of the cellular protein, signaling and transcriptional networks are in 'weak linkage' with each other [10]. A 'weak linker' is an interacting partner that binds with low affinity or only transiently to the other partner. This concept is used mainly in the context of networks and can refer both to physical interactions and influences of a network element on another network element. In metabolic networks, weak links are those reactions, which have a low flux [11-14]. In this paper we define weak linkers as small molecules and drugs that interact with cellular proteins using low-affinity physical interactions. Thus, most multi-target drugs are weak linkers. Because most links in cellular networks are weak, a low-affinity multi-target drug might be sufficient to achieve a significant modification. However, drug efficacy is also a highly important issue, and the question arises: can the multiplication of low-affinity binding affect the complex cellular system equally or better than the high-affinity and selective binding to a single target?

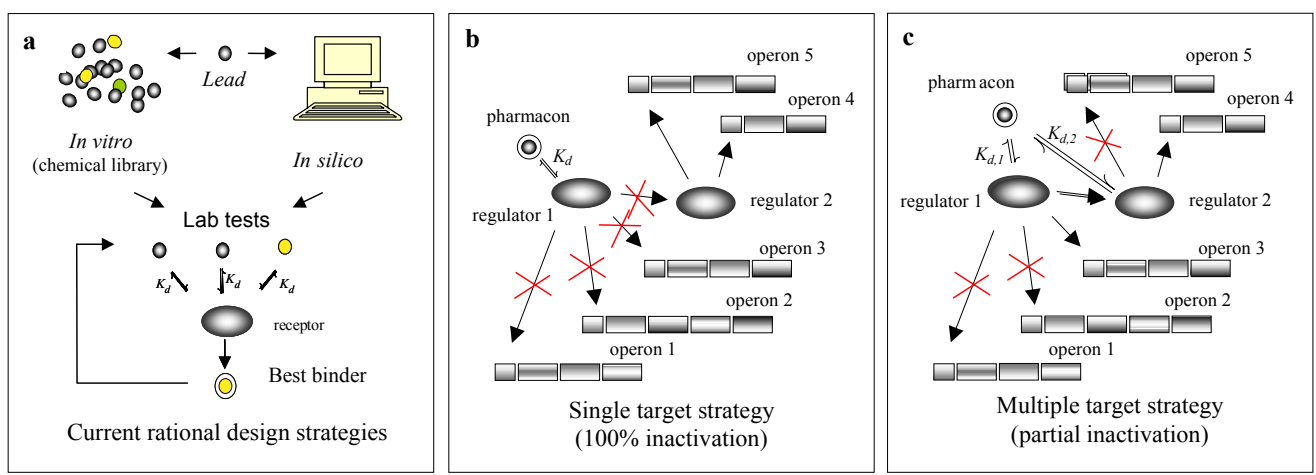

**Figure 1.** Drug design. **(a)** In the current single-target drug design paradigm the goal is to: (i) find a target gene, regulatory DNA sequence, protein or other macromolecule (such as non-coding RNA) of suitable function; (ii) identify the 'best-binder' by high-throughput screening of large combinatorial libraries (*in vitro*) and/or by rational drug design based on the three-dimensional structure of the target (*in silico*); (iii) provide a set of proof-of-principle experiments (lab tests); and (iv) develop a technology platform that predicts potential clinical applications. **(b,c)** In the network models, the drug candidate molecule (pharmacon) binds to its target, which is a part of a cellular network. The effect of a drug candidate on a prokaryotic genetic regulatory network that consists of regulator proteins (1 and 2) that affect the action of operons (DNA sequences coding at least one protein) is shown. However, the network approach to study the efficiency of drug action can also be applied to eukaryotic genetic regulatory networks, and metabolic or signaling networks. (b) The effect of complete inactivation of a single target (regulator 1), which is the usual outcome of the current single-target drug design paradigm, is shown. The effects of regulator 1 are inhibited completely. (c) The partial inactivation of multiple targets (regulators 1 and 2), which we propose as a novel drug design paradigm resulting in the development of efficient multi-target drugs, is shown. (The activation of a single target or multiple targets would essentially result in a similar, but reversed action.) In this case, only some of the effects of the regulators are inhibited. $K_{d,1}$ and $K_{d,2}$ (c) usually represent an interaction of lower affinity than that indicated by $K_d$. A method for drug-efficiency analysis using the network approaches (b,c) is described in **Box 1**.

**Box 1. Analysis of the effect of multi-target attacks in a network-based model**

In the network model of pharmacological actions, elements of the network represent various targets (proteins, RNA sequences or DNA sequences), whereas the links that connect them, represent their interactions within the cell. Here, the efficient drug-induced inhibition of a single target is modeled by the elimination of all interactions at the representing element (Figure Ia, complete knockout). Partial inactivation of a drug-target in the network-context can be modeled in two different ways: the drug either knocks out a proportion (e.g. half) of the interactions of a given protein [Figure Ib(i); partial knockout] or the drug attenuates all interactions of a protein [Figure Ib(ii); attenuation]. In the 'attenuation' experiments an attack on a link is modelled by depicting the respective connection (link) as 50% (dashed line) or 25% (dotted line) of the original. Finally, a distributed, system-wide attack can affect any protein-protein interaction (any link) within the network. Again, two simplified strategies can be used either knocking out [Figure Ic(i); distributed knockout] or attenuating [Figure Ic(ii); distributed attenuation] individual interactions (links) of the network. All of these described attacks correspond to inhibition scenarios, where functions are entirely or partially blocked in a manner similar to what happens when an antibiotic acts on a pathogen. The effect of a drug that restores the normal function of an inhibited receptor can be modeled by analogous steps carried out in reverse. Low-affinity, multi-target drugs might achieve the modification of network interactions better if they are allosteric inhibitors or activators. However, in case of low-affinity, weak interactions (which is the case of most interactions in the cell [11]) a competitive inhibition might also be efficient.

Attacks decrease network integrity and make the interactions between distant elements more complicated [37]. The corresponding network property can be captured by a computable quantity, called network efficiency (Figure Id), which is a global measure of network integrity related to the shortest path length among each pair of elements within the network. The network efficiency is expressed as the sum of the reciprocals of the shortest (directed) path lengths between all pairs of elements [Figure Id(i), where $N$ denotes the number of elements in the network] [38]. The path length values for the network shown in Figure Id(ii) are given by the $d_{ij}$ distance matrix [where the numbered interactions are shown in bold, and the numbers within the grid represent the number of steps in the interaction (e.g. '1' denotes a direct interaction between two elements, '2' denotes that two elements interact via two other intermediary elements)] [Figure Id(iii)]. Numbers shown in red are those paths that disappear when the two links marked in Figure Id(ii) are inhibited completely. In the case of attenuation, the path lengths concerned are multiplied by a weight of 2 or 4 depending on the extent of attenuation (50% or 25%) and the one with a minimum weight is selected [6,38]. If the two links marked by red X are deleted, the network efficiency decreases to 57.5%. The two major assumptions of the use of network efficiency as a measure for drug efficiency are that: (i) a mechanism targeted by a drug can be represented as a network; and (ii) all elements of this network must interact for the function of the targeted mechanism. Having more detailed information on the specific role and importance of the individual network elements, the efficiency of single-target and multi-target drugs can be compared more precisely. We give examples for these dynamic network models [20-22,25-33] in the text.

**Analysis of drug targets using a network approach**

Most studies that examine drug-development strategies are based on target-driven approaches, where an efficient method to combat a certain disease was sought. The network approach (Figure 1b-c) examines the effect of drugs in the context of a network of relevant protein-protein interactions [12-14]. In these network models each element represents a protein, and each link corresponds to an interaction between two proteins of the cell. In this model the efficient drug-induced inhibition of a single target means that the interactions around a given target are eliminated, whereas partial inhibition can be modeled as a partial knockout of the interactions of the target (Box 1). In addition to protein-interaction networks, regulatory, metabolic and signaling networks can be subjected to a similar analysis.

The network approach now has a tradition in drug target analysis. Comparison of transcriptional networks from various genomes helps to identify the function of novel proteins and thus increases the number of potential drug targets [15,16]. Proteomic analysis of protein-protein interactions might identify protein contact surfaces as novel sites of drug action [17] and neural networks help drug design [18,19]. Metabolic control analysis (or flux-balance analysis) uses a vast set of experimental data, and calculates all metabolic rates of the metabolic network assuming that the rates of reactions producing a metabolite must be equal with the rates of reactions, which consume it. These methods can highlight key points of the metabolism, where a parasite or a pathological metabolism can be targeted [20-22].

**Figure I.** Analysis of drug-induced target inhibition in the context of a network model.

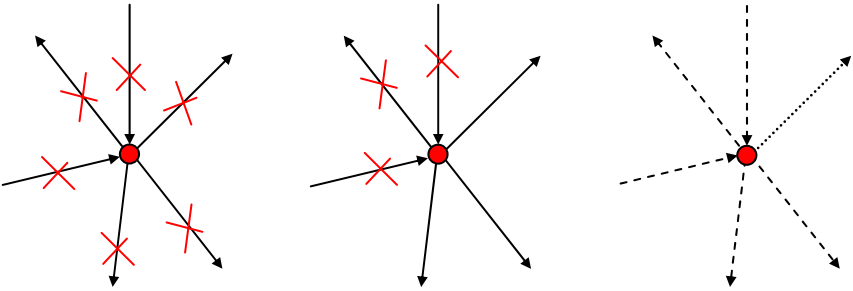

**(a)** Complete knockout      **(b)** Partial inactivation of several targets
                                    (i) Partial knockout      (ii) Attenuation

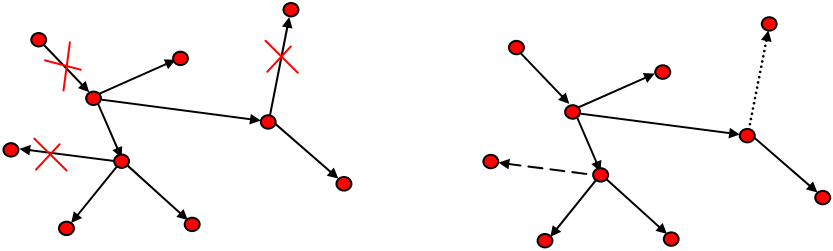

**(c)** Distributed system-wide attack
(i) Distributed knockout      (ii) Distributed attenuation

(i)
$$\textit{Efficiency} \quad E = \frac{\sum_{i \neq j} \frac{1}{d_{ij}}}{N(N-1)}$$

(iii) $d_{ij}$ matrix:

|    | 1 | 2 | 3 | 4 | 5 | 6 | 7 | 8 | 9 | 10 |
|----|---|---|---|---|---|---|---|---|---|----|
| 1  | - | 1 | 3 | 2 | 3 | 3 | 2 | 3 | 2 | 1  |
| 2  | - | - | - | - | - | - | - | - | - | -  |
| 3  | - | - | - | - | - | - | - | - | - | -  |
| 4  | - | 1 | 1 | - | 1 | - | - | - | - | -  |
| 5  | - | - | - | - | - | - | - | - | - | -  |
| 6  | - | - | - | - | - | - | - | - | - | -  |
| 7  | - | - | - | - | - | 1 | - | 1 | - | -  |
| 8  | - | - | - | - | - | - | - | - | - | -  |
| 9  | - | - | - | - | - | - | - | - | 1 | -  |
| 10 | - | 2 | 2 | 1 | 2 | 2 | 1 | 2 | 1 | -  |

E     = 0.181
$E_{xx}$ = 0.104 (57.5%)

(ii)

**(d)** Calculation of network efficiency

However, most of these methods have been used so far to steer target-identification attempts to single targets and a systematic analysis of multi-target drug action is still to come. Most of the above methods are, in principle, appropriate for the purpose; however, in most cases an adequate analysis appears to depend on too many parameters. However, simple topological network models might provide some preliminary insight (Box 1). In these simplified models, the 'attack' on a network, such as the genetic regulatory networks of *Escherichia coli* [23] or *Saccharomyces cerevisiae* [24], is modeled by removing or attenuating a target, which is either a protein (element), or an interaction (link). A comparison of various strategies suggests that multiple but partial attacks on carefully selected targets are almost inevitably more efficient than the knockout of a single, though equally well selected, target [6]. For example, the largest damage to the *E. coli* regulatory network is reached by removing an element with 72 connections. However, the same damage can be attained, if 3 to 5 nodes are partially inactivated [6]. A plausible explanation of this higher efficiency might be that even partial multi-target attacks block an increased number of individual interactions (network links) than a single knockout. Although this poses no problem from the practical point of multi-target drug action, simulations show that multiple attacks can be more efficient than a single attack even if the number of affected interactions is the same [6]. Thus, the reason underlying the efficiency of multi-target attacks is not trivial even from a theoretical point of view: multi-target attacks are not only better because they affect the network at more sites, they can, especially if distributed in the entire network, perturb complex systems more than concentrated attacks even if the number of targeted interactions is the same. Our initial analysis [6] (Box 1) was based on network topology, which can fit the case of, among others, antimicrobial drugs, where network damage corresponds well to the desired drug action. For the analysis of multi-target drugs that affect specific disease models (e.g. anti-hypertensive, anti-psychotic and anti-diabetic drugs), more specific signaling, metabolic and transcriptional network models are needed. However, the surprising generality of network behavior [11-14], in addition to the successful multi-target drugs mentioned earlier suggest that there are many highly efficient low-affinity, multi-target drugs awaiting discovery. The extension of current experimental [25-28] and modeling [20-22,29-33] approaches to perturb networks and mimic the effect of multi-target drugs would test the generality of our assumptions.

**Concluding remarks: towards a multi-target drug design paradigm**

We propose that drugs with multiple targets might have a better chance of affecting the complex equilibrium of whole cellular networks than drugs that act on a single target. Moreover, it is sufficient that these multi-target drugs affect their targets only partially, which corresponds well with the presumed low-affinity interactions of these drugs with several of their targets. Low-affinity, multi-target drugs might have another advantage. Weak links have been shown to stabilize complex networks, including macromolecular networks, ecosystems and social networks, buffering the changes after system perturbations [11,34]. If multi-target, low-affinity drugs inhibit their targets, they change a strong link into a weak link instead of eliminating the link completely. A weak activation also results in a weak link in most of the cases. Thus, multi-target drugs can increase the number of weak links in cellular networks and thus stabilize these networks in addition to having multiple effects.

How should we develop multi-target drugs? In recent years several experimental and modeling approaches have been developed to identify single targets in a network context [20-22,25-33]. Appropriate modifications of these approaches can provide several tools to help identify a suitable set of parallel targets and multi-target drug molecules for a particular disease. A high-throughput screen of the possible combinations can be a formidable task. However, the 'game theory approach' (where the multitude of possible equilibrium conditions is simplified by using pre-set rules of the 'game') might be fruitful to simplify the complex sets of equilibrium conditions (with the introduction of 'multi-target drug design games' [34]). Additionally, and perhaps most importantly, 'old fashioned' drug development might come back: if you want to know the response of a complex system, 'ask' the system (by testing drug candidates in complex *in vivo* tests)! And, although microarray techniques might be useful to follow multi-target drug strategies, *in vivo* pharmacology (i.e. whole-animal studies) might become important again [35]. Here the 'story' goes back to genetics: for more efficient *in vivo* testing, better animal models are needed. Better animal models can be achieved by 'humanizing' the metabolism and signaling of test animals. Disease target genes and their protein products might be transformed from drug targets to core elements of better animal models in the future.

The idea of multi-target attacks is not new. Perhaps the first, formal advocate of the multi-target approach was the military strategist Carl von Clausewitz who argued that instead of striving for successful single battles, strategy should simultaneously aim at 'the enemy's forces, his resources, and his will to fight' [36]. His complex approach proved to be an efficient antidote to Napoleon's rationally designed campaigns. Therefore, we think it is appropriate to conclude – paraphrasing another dictum of Clausewitz – that multi-target drugs and the network approach might become useful as the continuation of drug design by other means.

**Acknowledgements** We acknowledge the useful comments of Csaba Pál, Balázs Papp, Viktor Müller and Eörs Szathmáry (Eötvös Loránd University, Budapest, Hungary), Imre Boros, Péter Maróy and István Raskó (Szeged Biological Research Center, Hungary), Tamás Vicsek (Department of Biological Physics, Eötvös University, Budapest, Hungary), Andrew Young (Amylin Co., San Diego, CA USA), the anonymous referees and the Editor. Work in our laboratory was supported by research grants from the EU (FP6506850), Hungarian Science Foundation (OTKA-T37357, F-47281), Hungarian Ministry of Social Welfare (ETT-32/03), Hungarian Ministry of Economy (KKK-0015/3.0); Hungarian Office of Research and Development (OMFB-01887/2002, OMFB-00299/2002; NFKP-1A/056/2004) and EU-project ORIEL (IST-2001-32688) coordinated by the European Molecular Biology Organization (EMBO). S. P. is recipient of the Szent-Györgyi Award for teaching at the Department of Genetics and Molecular Biology, University of Szeged.